# Bond Relaxation and Electronic Properties of Two-Dimensional Sb/$MoSe_2$ and Sb/$MoTe_2$ Van der Waals Heterostructures

Maolin Bo,* Hanze Li, Zhongkai Huang, Lei Li, Chuang Yao

*Chongqing Key Laboratory of Extraordinary Bond Engineering and Advanced Materials Technology (EBEAM), Yangtze Normal University, Chongqing 408100, China*

**E-mail:* bmlwd@yznu.edu.cn

## Abstract

Van der Waals heterostructures have recently garnered interest for application in high-performance photovoltaic materials. Consequently, understanding the basic electronic characteristics of these heterostructures is important for their utilisation in optoelectronic devices. The electronic structures and bond relaxation of two-dimensional (2D) Sb/transition metal disulfides (TMDs, $MoSe_2$, and $MoTe_2$) van der Waals heterostructures were systematically studied using the bond-charge (BC) correlation and hybrid density functional theory. We found that the Sb/$MoSe_2$ and Sb/$MoTe_2$ heterostructures had indirect band gaps of 0.701 and 0.808 eV, respectively; further, these heterostructures effectively modulated the band gaps of $MoSe_2$ (1.463 eV) and $MoTe_2$ (1.173 eV). The BC correlation revealed four bonding and electronic contributions (electron-holes, antibonding, nonbonding, and bonding states) of the heterostructures. Our results provide an in-depth understanding of the Sb/TMD van der Waals heterojunction, which should be utilised to design 2D metal/semiconductor-based devices.

## 1. Introduction

Graphene has led to a research boom in the exploration of two-dimensional (2D) materials of various atomic layer thicknesses[1], such as transition metal disulfides (TMDs)[1], black phosphorus[2], graphitic-$C_3N_4$[3]. Due to their excellent optoelectronic properties, these materials have been extensively studied and their potential applications in next generation photovoltaic devices have been demonstrated. Because of its many advantages, including high thermal stability, high current carrying mobility, good thermal conductivity, and high conductivity[4, 5], a 2D material of Sb metal of the VA group element recently caused a huge sensation[6, 7]. These favourable characteristics make Sb monolayers a promising material for use in optoelectronic devices.[8]

With a focus on a single 2D material of TMDs, van der Waals (vdW) heterostructures, which are fabricated by vertically stacking different 2D semiconductors, open up new avenues for developing novel materials and devices.[9-12] In this type of heterostructure, strong covalent bonds within each monolayer ensure in-plane stability while relatively weak vdW interactions hold the stacked layers together and are expected to create new degrees of freedom in the electronic properties. This practice has been confirmed in GeS/$WS_2$[13] and $MoS_2$/AlN[14] heterostructures. Recently, TMDs and Sb, including Sb vdW heterostructures (e.g., Sb/InSe[15] and Sb/$WSe_2$[16]) are considered potential candidates for a high-performance photovoltaic materials due to their excellent semiconducting characteristics. On one hand, the lattices of Sb and TMD monolayers are perfectly matched, which means that constructing Sb/TMDs vdW heterostructures is highly feasible in practice. On the other hand, due to the high hole mobility in Sb and the high electron mobility in TMDs, combining Sb and TMD is expected to improve the transport properties of charge carriers in the Sb/TMDs vdW heterostructures. Therefore, exploring the basic electronic characteristics of Sb/TMDs vdW heterostructures is especially important if they are to be used as optoelectronic materials.

Compared to bulk materials, the surface of 2D materials has a considerable number of nonbonding electrons. The presence of these nonbonding electrons and their combinations affects the Hamiltonian of the solid and the associated physical and chemical properties of the 2D material. These early 2D metal/semiconductor heterojunctions that we studied are based on the bond-order–length–strength (BOLS) theory. A complete theoretical model for studying nonbonding electrons in 2D metal/semiconductor heterojunctions does not exist. Therefore, it is necessary to distinguish the influence of nonbonding and bonding electrons on the bonding properties of metal–semiconductor 2D heterojunctions; further it is important to regulate the electronic properties of the heterojunction according to the nature of the chemical bonding.

We investigated the electronic properties of 2D Sb/TMDs heterojunctions using BOLS theory together with density functional theory (DFT) calculations and the bond–charge (BC) model.[17] The electronic performance of Sb/TMDs arises from the intrinsic relationship between related antibonding, nonbonding andbonding , quantitatively characterising the role of crystal potential. Our research may lead to an increased interest in theoretical and experimental studies of Sb/TMDs vdW heterostructures of chemical bonding and open new avenues for future work in nanoelectronic heterostructures.

## 2. Methods

### 2.1 DFT calculations

All structural relaxation and electronic properties of $Sb/MoTe_2$ and $Sb/MoSe_2$ vdW heterostructures were calculated with CASTEP, which used DFT with a plane-wave pseudopotential.[18, 19] This was aimed at analysing the atomic structure, energetics, and electronic properties of 2D vdW heterojunctions. We used the HSE06[20] hybrid density function to describe the electron exchange and correlation potential; the cut-off energy of the plane-wave basis set was 440 eV. The k-point grids were $4\times5\times2$ ($Sb/MoTe_2$) and $4\times4\times1$($Sb/MoSe_2$), as shown in Table 1. The vacuum thickness was 18 Å. We chose structures with lattice strains less than 1% in QuantumWise, and the lattice parameters of each structure that were determined using QuantumWise are shown in Table 1. Moreover, the $Sb/MoTe_2$ and $Sb/MoSe_2$ vdW heterostructures were as shown in Fig. 1a and b. Additionally, in order to consider the long-range vdW interaction, we used Grimme's DFT dispersion correction; this was because the standard HSE06 function does not describe the weak interaction well.[21] In the calculations, the energy converged to $10^{-6}$ eV and the force on each atom converged to <0.01 eV/Å.

### 2.2 BC Model

In theory, we intended to extend the existing BOLS theory and combine it with DFT calculation to cover the local perturbation of the interfacial crystal potential and thereby overcome the nonbonding and antibonding electrons of other methods when dealing with electronic quantification of the heterojunction interface. The change in energy caused by external fields, such as pressure, temperature, electric fields, was used as a variable to determine the change in chemical bonds with the local electron density; thereby changing the crystal potential energy. The following formulae were used to describe the electronic state of a specific chemical bond:

$$\Delta E_v(i) = \left\langle v,i \middle| V_{cry}(r)(1+\Delta_H) \middle| v,i \right\rangle \left[ 1 + \frac{z\left\langle v,i \middle| V_{cry}(r)(1+\Delta_H) \middle| v,j \right\rangle}{\left\langle v,i \middle| V_{cry}(r)(1+\Delta_H) \middle| v,i \right\rangle} \right]$$

$$\cong E_B(1+\Delta_H)(1+\left(\frac{z\beta}{\alpha} < 3\%\right)) \cong E_i \quad (1)$$

$$\frac{E_i}{E_B} = \left(\frac{1/d_i}{1/d_B}\right)^m \propto \frac{\Delta E_v(i)}{\Delta E_v(B)} \propto \frac{V_{cry}(i)}{V_{cry}(B)} = \gamma \quad (2)$$

$$V_{cry}(B) = \gamma V_{cry}(i), \begin{cases} \gamma > 1, & \textit{deepening the potential well} \\ \gamma < 1, & \textit{strengthening the potential Energy Barrier} \end{cases} \quad (3)$$

$$\delta\rho = \frac{\rho(r_{ab}) - \rho(r_a) - \rho(r_b)}{\rho(r_{cd}) - \rho(r_c) - \rho(r_d)} \propto \frac{\Delta E_v(ab) / \iiint d^3 r_{ab}}{\Delta E_v(cd) / \iiint d^3 r_{cd}} \propto \frac{E_{ab} / r_{ab}^3}{E_{cd} / r_{cd}^3} \quad (4)$$

$$V_{cry}(i) = (\frac{1}{4\pi\varepsilon_0}) \iiint d^3 r_i \iiint d^3 r_i \frac{\Delta\rho^{\text{hole}}(r_i)\Delta\rho^{\text{electron}}(r_i)}{r_i} \quad (5)$$

Here, $d_i$ is the bond length of the atom; $E_i$ represents the single bond energy, the subscript $b$ represents bulk atoms, $\Delta E_v(i)$ is the electronic binding energy shift, $V_{cry}(i)$ is the crystal potential, $\Delta\rho(r_i)$ is the deformation charge density, $r_i$ is the radius of the atom, and $m$ is the bond nature indicator. **Eq. 1** establishes the relationship between electronic binding energy and bond energy from the energy band theory.[22] **Eq. 2** explains the relationship between chemical bonds (energy and length) and crystal potential functions based on the BOLS theory.[17] In **Eq. 3**, $\gamma V_{cry}$ ® may become deeper ($\gamma >$ 1 for a potential well formation) or shallower ($\gamma <$ 1 for a potential barrier formation) than the corresponding $V$®($r$) of the specific constituent.[23] **Eq. 4** describes the relationship between electronic binding energy and deformation charge density, and **Eq. 5** shows the relationship between the crystal potential and the deformation charge density. From the above model, we obtained the functional relationship between bond-energy–electronic binding–energy–deformation-charge density and crystal potential energy through dimensional analysis and unit conversion. **Fig. 2** shows a schematic of the BC model.

## 3. Results and discussion

### 3.1 Formation energy and structures

The electronic structures and band gaps of Sb, $MoTe_2$, $MoSe_2$, Sb/$MoTe_2$, and Sb/$MoSe_2$ were calculated. We considered the influence of lattice strain and vdW forces for the Sb/$MoTe_2$ and Sb/$MoSe_2$ heterostructures. The lattice strains in the Sb/$MoTe_2$ and Sb/$MoSe_2$ heterostructures were 0.52 % and 0.49 %, respectively, which was less than 1 % in both cases. The 2D heterostructures require spacing to maintain their stability, as shown in **Fig. 2**. The relaxed lattice parameters and interlayer distances are listed in **Table 1**. We set the initial distance in the *z* direction between the top Sb metallic layer and the bottom TMD layer to 3.20 Å; moreover, the relaxed interlayer distances in the Sb/$MoTe_2$ and Sb/$MoSe_2$ heterostructures were 3.94 and 3.75 Å, respectively.

We also calculated the formation energies ($E_{form}$) of the Sb/$MoTe_2$ and Sb/$MoSe_2$ heterostructures using the following equation[14]:

$$E_{form} = E_{total}^{heterostructure} - E_{total}^{MoSe_2(MoTe_2)} - E_{total}^{Sb} \text{ (eV)}. \quad (6)$$

$E_{form}$ of the Sb/$MoTe_2$ and Sb/$MoSe_2$ heterostructures were -0.37 and 0.21 eV, respectively. These negative formation energies indicate that the 2D heterostructure were structurally stable.

### 3.2 Band structure and density of states

In order to assess the validity of our calculation, we calculated the band structure of the monolayer Sb, $MoTe_2$ and $MoSe_2$ structures, as illustrated in **Fig. 3**. The calculated band gaps of $MoTe_2$ and $MoSe_2$ were 1.173 eV and 1.463 eV, respectively; this was consistent with previously reported values[24]. The monolayer Sb($MoTe_2$) and Sb($MoSe_2$) had indirect band gaps that were calculated to be 0.860 eV and 0.801 eV, respectively. However, the electronic contributions of the band gaps of the Sb/$MoTe_2$ and Sb/$MoSe_2$ structures were 0.701 eV and 0.808 eV, respectively, which shows that the Fermi level($E_f = 0$) is primarily determined by Sb.

The local densities of states (LDOS) of the Sb/$MoTe_2$ and Sb/$MoSe_2$ heterostructures are shown in **Fig. 4**. The *s*-, *p*-, and *d*- orbital contributions to the LDOS of $MoTe_2$ and $MoSe_2$ at the Fermi level are shown in **Figs 4a** and **4b**. **Fig. 4c** and **4d** show the Fermi-level LDOS contributions from Sb, Mo and Se atoms. We compared the LDOS values of the Sb/$MoTe_2$ and Sb/$MoSe_2$ heterostructures. For Sb/$MoSe_2$, the top of the valence band was mainly determined by Sb, while the bottom of the conduction band was mainly determined by the mixing of Sb, Mo, and Se. The valence and conduction bands of Sb/$MoTe_2$ were equally influenced by Sb. The Sb-based heterostructures modulated the band gap of the $MoTe_2$ and $MoSe_2$. Moreover, both the Sb/$MoTe_2$ and the Sb/$MoSe_2$ heterostructure showed indirect band gaps (0.701 eV and 0.808 eV), that were suitable for visible-light absorption.

### 3.3 Deformation charge density resolved bond and electrons features

We believe that the formation of Sb/$MoTe_2$ and Sb/$MoSe_2$ vdW heterostructures is primarily due to bond, electron, and charge contributions. The deformation charge density is calculated by DFT and can be obtained via contribution from four bonding and electronic features of bonding states, nonbonding states, electron holes, and antibonding states; the calculated value is consistent with the bond–band-barrier predictions[25]. The deformation charge densities of Sb/$MoTe_2$ and Sb/$MoSe_2$ are shown in **Fig. 5**. The deformation charge density scale indicates the charge value.

For the Sb/$MoTe_2$ and Sb/$MoSe_2$ heterostructures, the increase and decrease in the number of electrons are shown in blue and red, respectively. $\Delta\rho(r_i) > 0$ causes the electron density to increase and electrons are obtained from atoms in bonding or nonbonding states (shown in red). Conversely, $\Delta\rho(r_i) < 0$ results in a decrease in electron density and electrons are lost from atoms in antibonding states or that have electron holes (shown in blue). Using **Eq. 5**, we calculated crystal potentials of the bonding, nonbonding, and antibonding states of heterostructures, and the results are shown in **Table 3**. These findings will be useful for calculation ing the bond states and crystal potentials of 2D vdW heterostructures.

## 4. Conclusions

Combining the BC correlation with DFT calculations has provided insight into the physical origin of bond relaxation and the electronic properties of $Sb/MoTe_2$ and $Sb/MoSe_2$. The BC correlation revealed four bonding contributions to the deformation charge densities, namely: nonbonding electrons, antibonding electrons, electron holes and bonding electrons. The band gaps of $Sb/MoTe_2$ and $Sb/MoSe_2$ were calculated to be 0.701 eV and 0.808 eV, respectively. Overall, our theoretical predictions indicate that $Sb/MoTe_2$ and $Sb/MoSe_2$ are potential materials for application in 2D metal–semiconductors. These findings will be useful in the design of optoelectronic devices.

**Acknowledgment:**

Financial support was provided by Program of Chongqing Municipal Education Commission (Grant No. KJQN201901421)

**Figures and Tables**

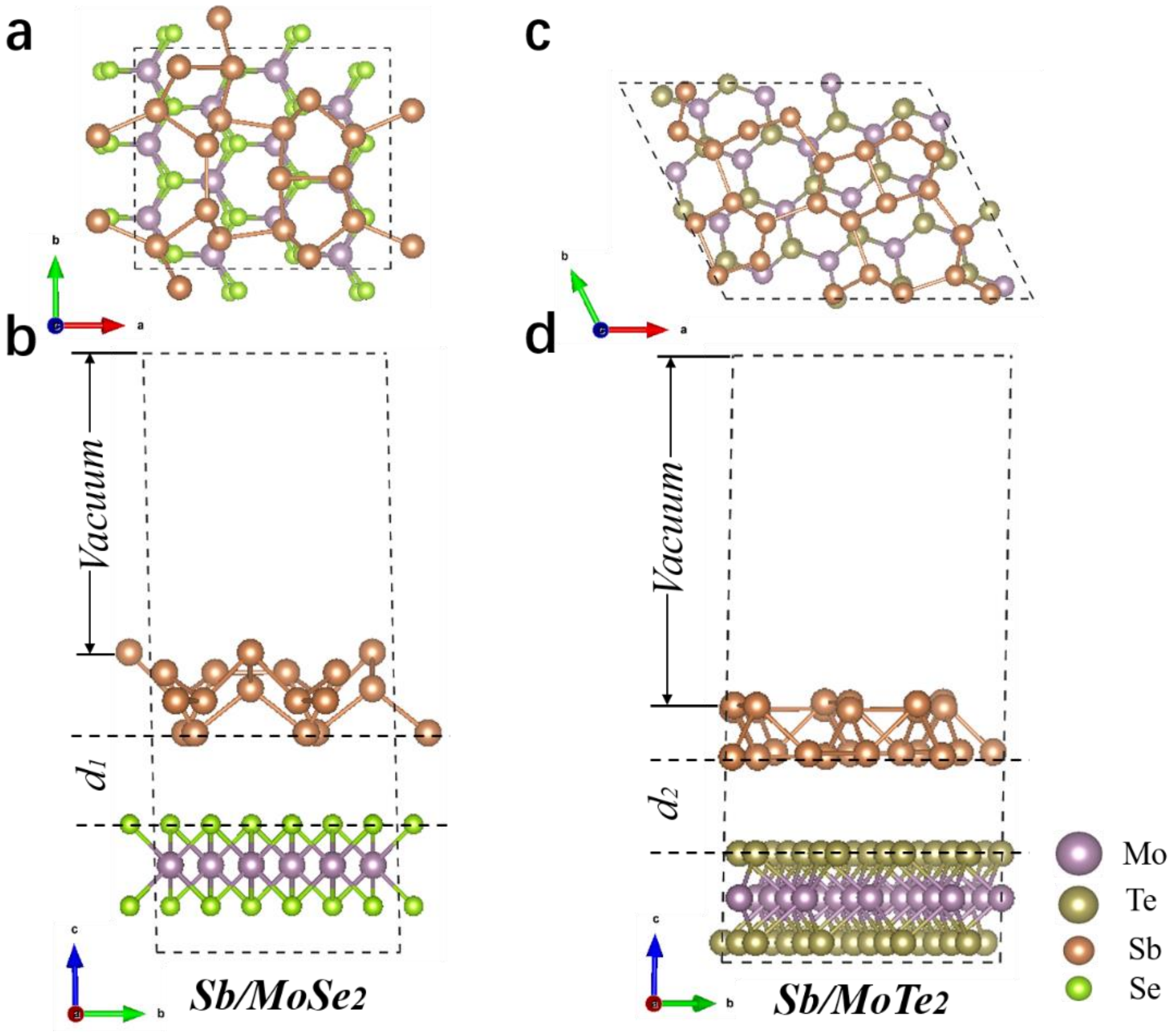

**Fig. 1** (a) and (b) are top views of the Sb/$MoSe_2$ heterostructure. (c) and (d) are side views of the Sb/$MoTe_2$ heterostructure. The interlayer distance of Sb/$MoSe_2$ $d_1$ = 3.75 Å and Sb/$MoTe_2$ $d_2$ =3.94 Å.

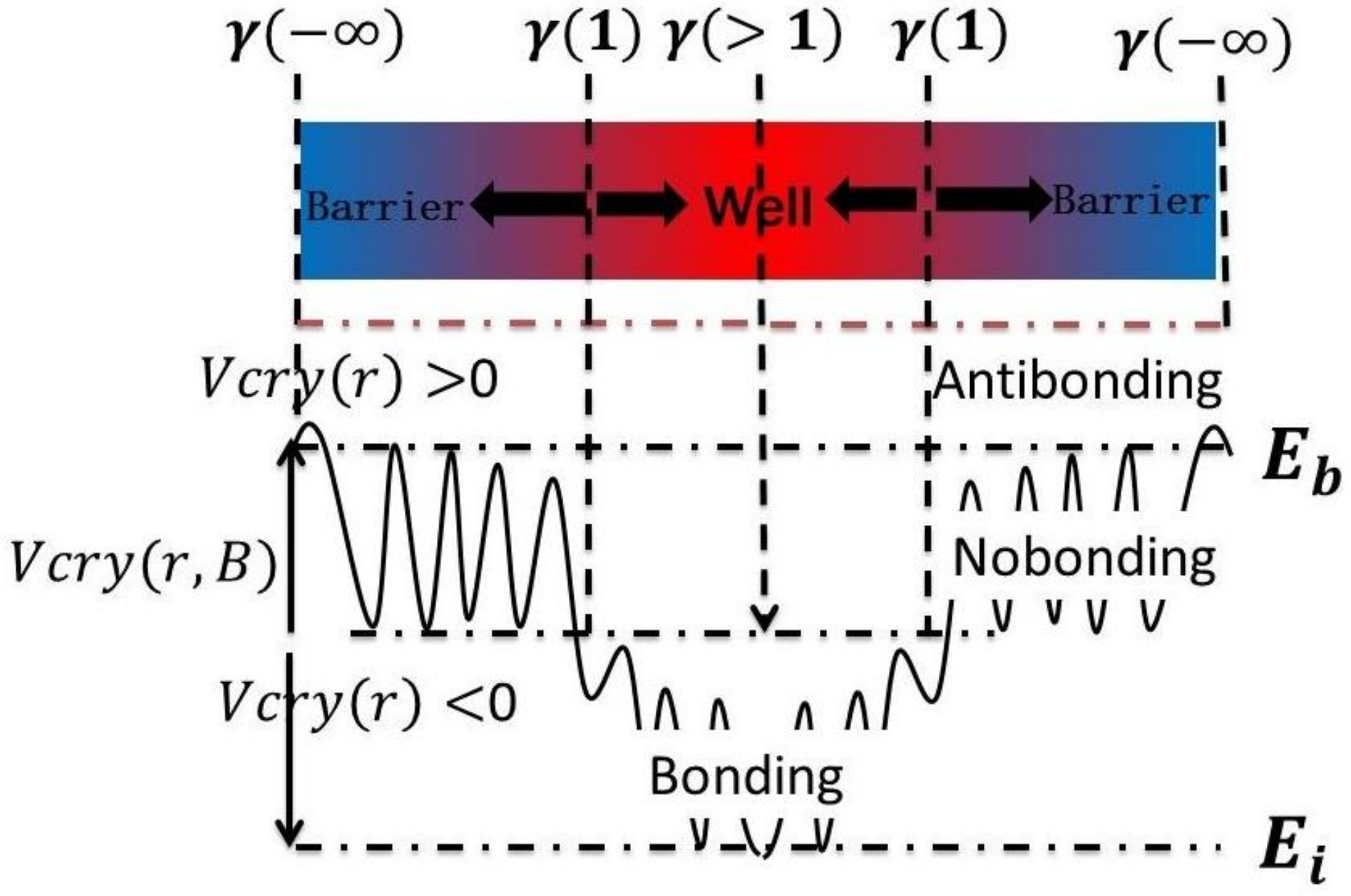

**Fig. 2** Schematic of the BC correlation.

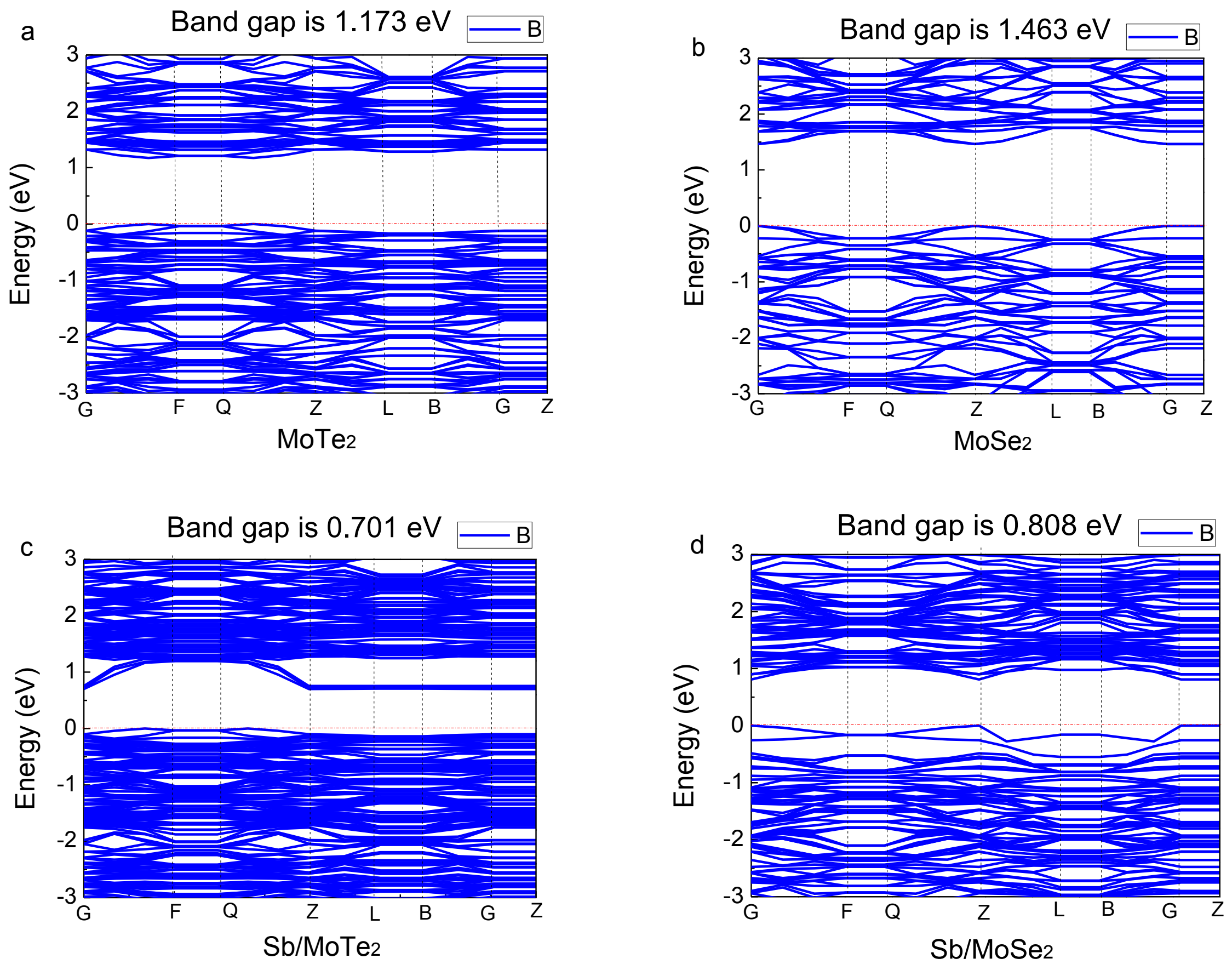

**Fig. 3** Band structure of (a) $MoTe_2$, (b) $MoSe_2$, (c) Sb/$MoTe_2$, and (d) Sb/$MoSe_2$.

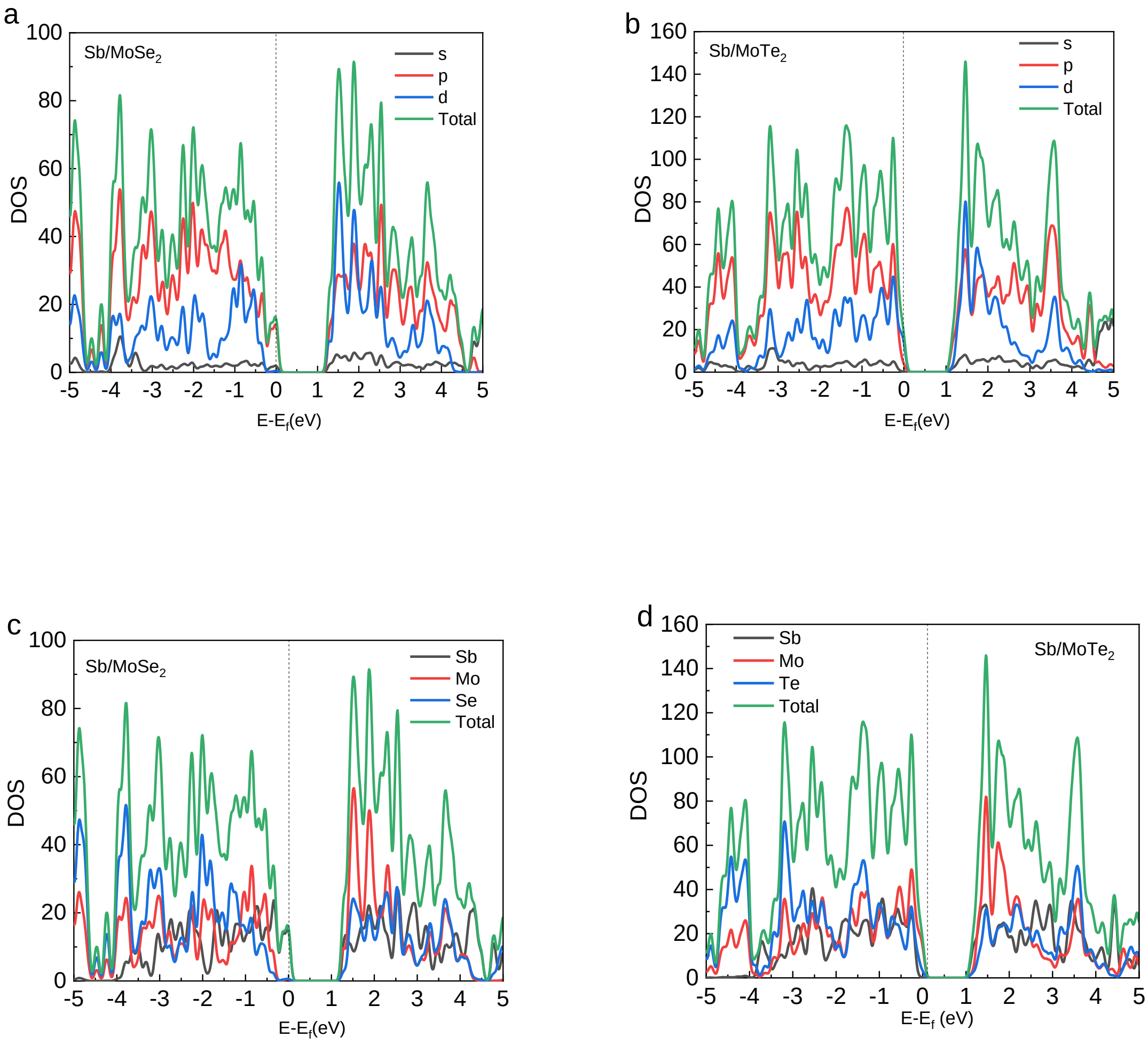

**Fig. 4** LDOS of Sb/MoTe$_2$ and Sb/MoSe$_2$ heterostructures.

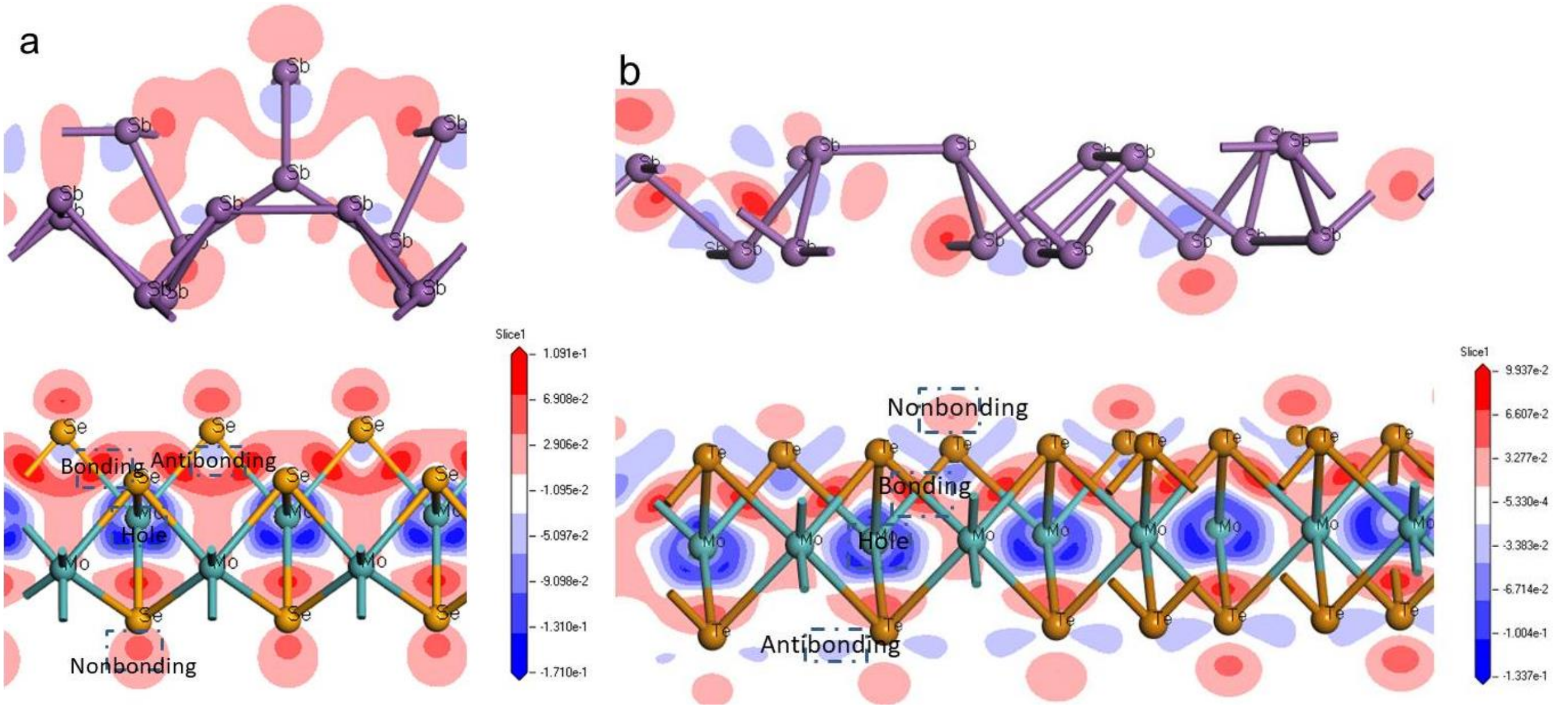

**Fig. 5** Deformation charge density resolved bond and electron features: antibonding, nonbonding and bonding states as well as electron holes.

**Table 1**. (a) angles and lattice parameters of each heterostructure. (b) k-point grids, bandgap, layer-spacing, and lattice strain (ε) of each heterostructure.

(a)

| | Angles | | | Lattice parameters | | |
|---|---|---|---|---|---|---|
| Heterostructure | α | β | γ | a | b | c |
| Sb/$MoTe_2$ | 89.66 ° | 88.69 ° | 116.43 ° | 16.23Å | 12.69Å | 24.72Å |
| Sb/$MoSe_2$ | 91.30 ° | 96.46 ° | 90.00 ° | 11.58Å | 9.94Å | 24.57Å |

(b)

| | k-points | Band gap (eV) | layer spacing (Å) | lattice strain ε (%) |
|---|---|---|---|---|
| Sb/$MoTe_2$ | 4 × 4 × 1 | 0.701 | 3.94 | 0.52 |
| Sb/$MoSe_2$ | 4 × 5 × 2 | 0.808 | 3.75 | 0.49 |

**Table 2** The formation energy of Sb/$MoTe_2$ and Sb/$MoSe_2$ heterostructures.

$$E_{form} = E_{total}^{heterostructure} - E_{total}^{MoSe_2(MoTe_2)} - E_{total}^{Sb}$$

| $E_{total}^{heterostructure}$ (eV) | $E_{total}^{MoSe_2(MoTe_2)}$ (eV) | $E_{total}^{Sb}$ (eV) | $E_{form}$ (eV) |
|---|---|---|---|

| | | | | | | |
|---|---|---|---|---|---|---|
| Sb/$MoSe_2$ | -11250.19 | $MoSe_2$ | -8832.96 | Sb | -2417.02 | -0.21 |
| Sb/$MoTe_2$ | -14903.71 | $MoTe_2$ | -11277.93 | Sb | -3625.41 | -0.37 |

**Table 3** This table shows the deformed charge density difference ratio $\Delta\rho(r)$ and $V_{cry}(r)$ of the vdW heterojunctions as calculated using the BC correlation( $\varepsilon_0$ =8.85 × $10^{-12}C^2$ •$N^{-1}$•$m^{-2}$,e=1.60×$10^{-19}$C).

| | Sb/$MoTe_2$ ($r$ =$d$/2=2.725/2 Å) | Sb/$MoSe_2$ ($r$ =$d$/2=2.541/2 Å) |
|---|---|---|
| $\Delta\rho^{\text{hole}}(r)$ (e/$Å^3$) | $-1.337\times10^{-1}$ | $-1.710\times10^{-1}$ |
| $\Delta\rho^{bonding-electron}(r)$ (e/$Å^3$) | $9.937\times10^{-2}$ | $1.091\times10^{-1}$ |
| $\Delta\rho^{\text{nonbonding-electron}}(r)$ (e/$Å^3$) | $3.277\times10^{-2}$ | $6.908\times10^{-2}$ |
| $\Delta\rho^{\text{antibonding-electron}}(r)$ (e/$Å^3$) | $-3.383\times10^{-2}$ | $-5.097\times10^{-2}$ |
| $V_{cry}^{nonbonding}(r)$ (eV) | -0.296 | -0.562 |
| $V_{cry}^{bonding}(r)$ (eV) | -0.897 | -0.888 |
| $V_{cry}^{Antibonding}(r)$ (eV) | 0.305 | 0.415 |